\documentclass[
 reprint,
 amsmath,amssymb,superscriptaddress,
 aps,
 prl,
groupedaddress
]{revtex4-1}

\usepackage{graphicx}
\usepackage{dcolumn}
\usepackage{bm}
\usepackage{color}
\usepackage{ulem}
\usepackage{float}
\usepackage{textcomp}
\restylefloat{table}
\usepackage{verbatim}
\usepackage{epstopdf}

\begin{document}

\title{Geometric Resonance of Composite Fermions near Bilayer Quantum Hall States}
\date{\today}

\author{M. A.\ Mueed}
\author{D.\ Kamburov}
\author{L. N.\ Pfeiffer}
\author{K. W.\ West}
\author{K. W.\ Baldwin}
\author{M.\ Shayegan}
\affiliation{Department of Electrical Engineering, Princeton University, Princeton, New Jersey 08544, USA}

\begin{abstract}
Via the application of parallel magnetic field, we induce a single-layer to bilayer transition in two-dimensional electron systems confined to wide GaAs quantum wells, and study the geometric resonance of composite fermions (CFs) with a periodic density modulation in our samples. The measurements reveal that CFs exist close to bilayer quantum Hall states, formed at Landau level filling factors $\nu=1$ and 1/2. Near $\nu=1$, the geometric resonance features are consistent with half the total electron density in the bilayer system, implying that CFs prefer to stay in separate layers and exhibit a two-component behavior. In contrast, close to $\nu=1/2$, CFs appear single-layer-like (single-component) as their resonance features correspond to the total density.    
\end{abstract} 

\maketitle
A clean two-dimensional electron system (2DES), at low temperature and under perpendicular magnetic field ($B_{\perp}$) which quenches the kinetic energy into the discrete Landau levels, displays many exciting phases as a function of Landau level filling factor $\nu$ (number of electrons per flux quantum). Examples include the integer and fractional quantum Hall states (IQHS and FQHS), Fermi sea of composite fermions (CFs), spin texture Skyrmions, stripe and bubble phases, and Wigner crystal \cite{Jain.2007, Shayegan.Flatland.Review.2006}. The introduction of a ``layer" component adds to the rich physics as inter- and intra-layer interactions lead to yet more phases such as \textit{bilayer} Wigner crystal \cite{Manoharan.PRL.1996,Hatke.Ncomm.2015} as well as QHSs at total filling factor $\nu=1$ and 1/2 which are believed to be the two-component $\Psi_{111}$ and $\Psi_{331}$ states, respectively \cite{Lay.PRB.1997,Hasdemir.PRB.2015,Greiter.PRB.1992,Biddle.PRB.2013,Storni.PRL.2010,Spielman.PRL.2000,Eisenstein.ARCMP.2014,Kellogg.PRL.2004,Tutuc.PRL.2004,Peterson.PRB.2010,Thiebaut.PRB.2015,Mueed1.PRL.2015,Liu.PRL.2014,Liu.PRB.2014,Liu.PRB.2015,Moon.PRB.1995,Girvin.Book,Halperin.HPA.1983,Suen.PRL.1994,Lay.PRB.1994,Suen.PRL.1992,Murphy.PRL.1994,Eisenstein.PRL.1992,Shabani.PRB.2013}. Here, $\Psi_{mm'n}\sim{\displaystyle \prod_{i,j} (z_i-z_j)}^m{\displaystyle \prod_{i,j} (w_i-w_j)}^{m'}{\displaystyle \prod_{i,j} (z_i-w_j)}^{n}$ is the two-component Laughlin wave function \cite{footnoteA,footnoteC}. In this generalized description, $z$ and $w$ correspond to the complex coordinates of an electron in the different layers while the exponents $(m,m',n)$ characterize the wave function. These bilayer QHSs continue to be widely studied thanks to their many remarkable properties. In particular, the $\nu=1$ $\Psi_{111}$ QHS phase is generally understood to be a Bose-Einstein condensate of excitons (interlayer pairing of electrons and holes) which exhibits superfluid transport and interlayer tunneling behavior, similar to the d.c. Josephson effect \cite{Spielman.PRL.2000,Eisenstein.ARCMP.2014,Kellogg.PRL.2004,Tutuc.PRL.2004}. As for the $\nu=1/2$ QHS observed in wide single quantum wells, although generally considered to be the $\Psi_{331}$ state, there is still a debate on whether it might have a single-component (Pfaffian) origin \cite{Peterson.PRB.2010,Thiebaut.PRB.2015,Greiter.PRB.1992,Biddle.PRB.2013,Storni.PRL.2010,Suen.PRL.1994,Suen.PRL.1992,Shabani.PRB.2013,Liu.PRL.2014,Mueed1.PRL.2015,Liu.PRB.2014}.

Here we probe the bilayer QHSs at $\nu=1$ and 1/2 using the Fermi gas properties of CFs at nearby fillings. CFs are exotic electron-flux quasi-particles near $\nu=1/2$ in single-layer 2DESs that elegantly describe the physics of strongly interacting electrons \cite{Shayegan.Flatland.Review.2006,Jain.2007, Jain.PRL.1989,Halperin.PRB.1993}. Recently, CFs, and in particular whether or not they are particle-hole symmetric, have received intense attention \cite{Kamburov.PRL.2014, Kachru.PRB.2015,Geraedts.Science.2016,Wang.PRB.2016,Barkeshli.PRB.2015,Balram.PRL.2015, Son.PRX.2015,Balram.PRB.2016,Mulligan.PRB.2016}. In the CF picture, each electron combines with two flux-quanta. As a result, CFs feel $zero$ net $B_{\perp}$ at $\nu=1/2$. Away from $\nu=1/2$, however, they are subjected to the effective magnetic field $B_{\perp}^{*}=B_{\perp}-B_{\perp,1/2}$ \cite{footnote1} where $B_{\perp, 1/2}$ is the field at $\nu=1/2$. Thanks to this reduction in flux density, the neighboring FQHSs can be interpreted as the IQHSs of CFs \cite{Jain.PRL.1989,Jain.2007}. The flux attachment also explains the compressible state observed at $\nu=1/2$ in terms of a CF Fermi gas \cite{Halperin.PRB.1993}. This Fermi gas state extends to the vicinity of $\nu=1/2$ much like that of electrons near $B_{\perp}=0$. Many experimental studies have confirmed the existence of a CF Fermi sea near $\nu=1/2$ via the geometric resonance (GR) of CFs' cyclotron orbit with a periodic modulation of the 2DES density \cite{Willett.PRL.1993, Kang.PRL.1993, Goldman.PRL.1994, Smet.PRL.1996, Smet.PRL.1999, Willett.PRL.1999, Kamburov.PRL.2012, Kamburov.PRL.2013, Kamburov.PRB.2014, Kamburov.PRL.2014}. 

According to a recent study, signatures of CFs are also seen near total filling $\nu=1/2$ in $bilayer$ samples exhibiting a QHS at $\nu=1/2$ \cite{Mueed1.PRL.2015}. Surprisingly, the observed GR features correspond to the $total$ density in the bilayer system, suggesting a single-layer (or single-component) behavior for CFs \cite{Mueed1.PRL.2015}. This is particularly puzzling, if the nearby $\nu=1/2$ QHS is indeed the two-component ($\Psi_{331}$) state. Here, we examine the fate of CFs near another bilayer QHS, namely the QHS at $\nu=1$ which is also believed to be a two-component ($\Psi_{111}$) state. In sharp contrast to the $\nu=1/2$ case, our GR features are consistent with \textit{half} of the total density, revealing that CFs split into two layers near the bilayer QHS at $\nu=1$.

\begin{figure}
\includegraphics[width=.47\textwidth]{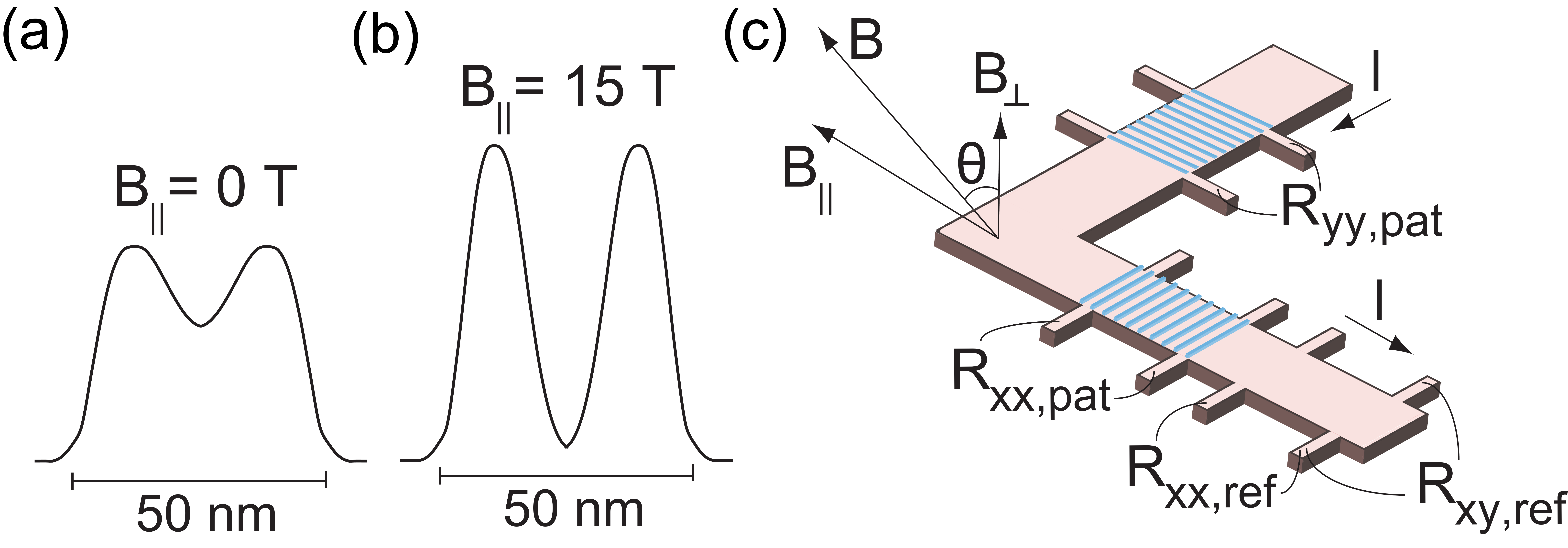}
\caption{\label{fig:Fig10} (color online) Self-consistently calculated charge distribution of a 2DES confined to a 50-nm-wide QW with density $n=1.7\times10^{11}$ cm$^{-2}$ for (a) $B_{||}=0$ and (b) 15 T. 
In both cases, we assume that $B_{\perp}=0$ T. (c) Schematics of the L-shaped Hall bar used in our study. The blue lines represent the surface superlattice. The sample is tilted in field at angle $\theta$ to introduce an in-plane field component ($B_{||}$). The sample's geometry allows measurement of magnetoresistance along ($R_{xx}$) and perpendicular ($R_{yy}$) to $B_{||}$.}
\end{figure}

In our measurements we apply a large parallel magnetic field ($B_{||}$) to tune the charge distribution and tunneling in our samples (Fig. \ref{fig:Fig10}). Without any $B_{||}$, electrons confined in a wide quantum well (QW) such as ours can be viewed as essentially a single-layer-like 2DES, or as two strongly coupled 2DESs with abundant interlayer tunneling (Fig. \ref{fig:Fig10}(a)). By tilting the sample at angle $\theta$ (as shown in (Fig. \ref{fig:Fig10}(c)), we apply $B_{||}$ which couples to the electron system through its finite layer thickness and adds an extra out-of-plane confinement, leading to a bilayer charge distribution with significantly suppressed interlayer tunneling (Fig. \ref{fig:Fig10}(b)) \cite{Hasdemir.PRB.2015,Mueed.PRL.2015,Lay.PRB.1997,Manoharan.PRL.1997}. 
Therefore, by changing $\theta$ and adjusting both $B_{||}$ and $B_{\perp}$, one can study the evolution of the 2DES from a single-layer to a bilayer and the stability of its phases at various $\nu$ \cite{Footnote.1}. In this work, we probe the QHSs and CFs near $\nu=1$ as well as $\nu=1/2$ as the system makes its single-layer to a bilayer transition at sufficiently large $\theta$.
Our sample, grown via molecular beam epitaxy, is a
50-nm-wide GaAs (001) QW located 190 nm
under the surface. The QW is flanked on each side by
150-nm-thick Al$_{0.24}$Ga$_{0.76}$As spacer and Si $\delta$-doped
layers. The low-temperature 2DES density is $n=1.71\times10^{11}$ cm$^{-2}$, and
the mobility is $\simeq10^{7}$ cm$^{2}/$Vs. We partially pattern the surface of the sample, an L-shaped Hall bar (Fig. \ref{fig:Fig10}(c)), by fabricating a strain-inducing
superlattice of 200 nm period.
The superlattice, made of negative electron-beam resist, imparts a density modulation of the same period to the 2DES through the piezoelectric effect in GaAs \cite{Endo.PRB.2005,Kamburov.PRB.2012,Skuras.APL.1997}. We pass current perpendicular to the density modulation (Fig. \ref{fig:Fig10}(c)). Whenever the CFs' cyclotron orbit diameter ($2R_{c}^{*}$) becomes commensurate with the modulation period ($a$), a GR should manifest in the magnetoresistance as a minimum. Since $2R_{c}^{*}=2{\hbar}k_{F}^{*}/eB_{\perp}^{*}$, the $B_{\perp}^{*}$-positions of the resistance minima directly yield the CFs' Fermi wave vector $k_F^{*}$. For a fully spin-polarized, circular Fermi contour, the expected positions of these minima are given by the condition $2R_{c}^{*}/a=i+1/4$ \cite{Smet.PRL.1999, Willett.PRL.1999, Kamburov.PRL.2012, Kamburov.PRL.2013, Kamburov.PRB.2014, Kamburov.PRL.2014}, where $i$ is an integer, $k_{F}^{*}={\sqrt{4{\pi}n^{*}}}$, and $n^{*}$ is the CF density. 
Throughout this manuscript, $R_{xx,pat}$ and $R_{yy,pat}$ refer to the longitudinal magnetoresistance of the patterned regions located on perpendicular arms of our Hall bar sample, and $R_{xx,ref}$ and $R_{xy,ref}$ represent the reference (unpatterned) region's longitudinal and transverse (Hall) magnetoresistances, respectively (Fig. \ref{fig:Fig10}(c)).

\begin{figure}
\includegraphics[width=.47\textwidth]{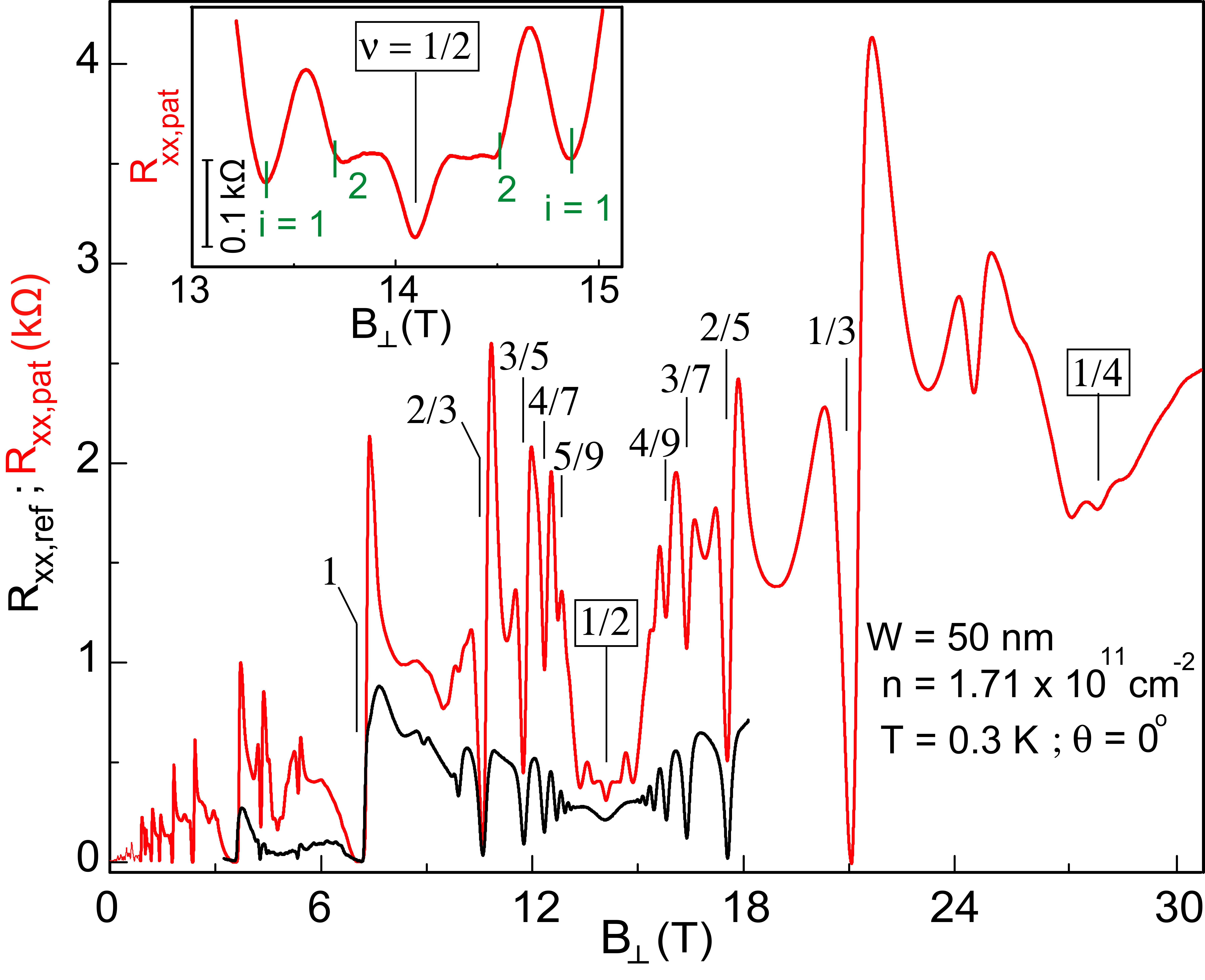}
\caption{\label{fig:Fig1} (color online) (a) $R_{xx,ref}$ (black) and $R_{xx,pat}$ (red) traces, measured at $\theta=0^{\circ}$, i.e., $B_{||}=0$. There are clear signatures of geometric resonance of CFs near $\nu=1/2$ and 1/4 in the $R_{xx,pat}$ trace. Inset: Magnified $R_{xx,pat}$ trace near $\nu=1/2$. The green lines mark the expected positions of the $i=1$ CF geometric resonance features (see text).}
\end{figure}


First, we present $R_{xx,ref}$ and $R_{xx,pat}$ traces taken at $B_{||}=0$ (Fig. \ref{fig:Fig1}). The traces resemble those of a standard, single-layer 2DES \cite{Kamburov.PRB.2014}. Near $\nu=1/2$, the $R_{xx,pat}$ trace exhibits features that are absent in the $R_{xx,ref}$ trace. These features are shown magnified in the inset where a V-shaped resistance dip is followed by one or more resistance minima, characteristic of the GR phenomena. We mark with green lines the expected positions of the $i=1$ and 2 resistance minima according to the commensurability condition mentioned earlier \cite{footnote1_1}. 
The resistance minima closely follow their expected positions based on a single-layer 2DES, evincing that at $B_{||}=0$ our 2DES is indeed single-layer-like. Note that the $i=1$ and 2 assignments of the GR features in Fig. \ref{fig:Fig1} and other cases in the manuscript can be verified by the ratio of their respective $B_{\perp}^{*}$-positions. For example, such ratio in Fig. \ref{fig:Fig1} inset is 1.9 which is very close to the expected value of $B_{\perp,i=1}^{*}/B_{\perp,i=2}^{*}$ (= 2.25/1.25 = 1.80) confirming that this assignment is indeed correct. 

\begin{figure*}
\includegraphics[width=.96\textwidth]{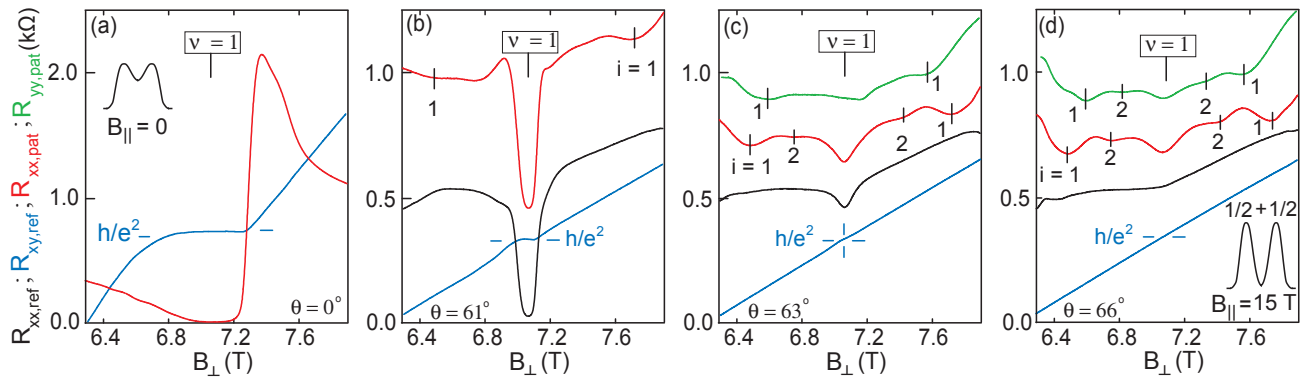}
\caption{\label{fig:Fig2} (color online) (a)-(d) Magnetoresistance traces near $\nu=1$ at different $\theta$. The $R_{xx,ref}$ and $R_{xy,ref}$ traces are shown in black and blue, respectively. The $R_{xx,pat}$ (red) and $R_{yy,pat}$ (green) traces correspond to current passing along and perpendicular to $B_{||}$ (Fig. \ref{fig:Fig10}(c)). For clarity, the traces are offset vertically. 
We also include the self-consistently calculated (assuming $B_{\perp}=0$ T) charge distributions for $B_{||}=0$ and 15 T in the insets of (a) and (d).}
\end{figure*}

\begin{figure}
\includegraphics[width=.32\textwidth]{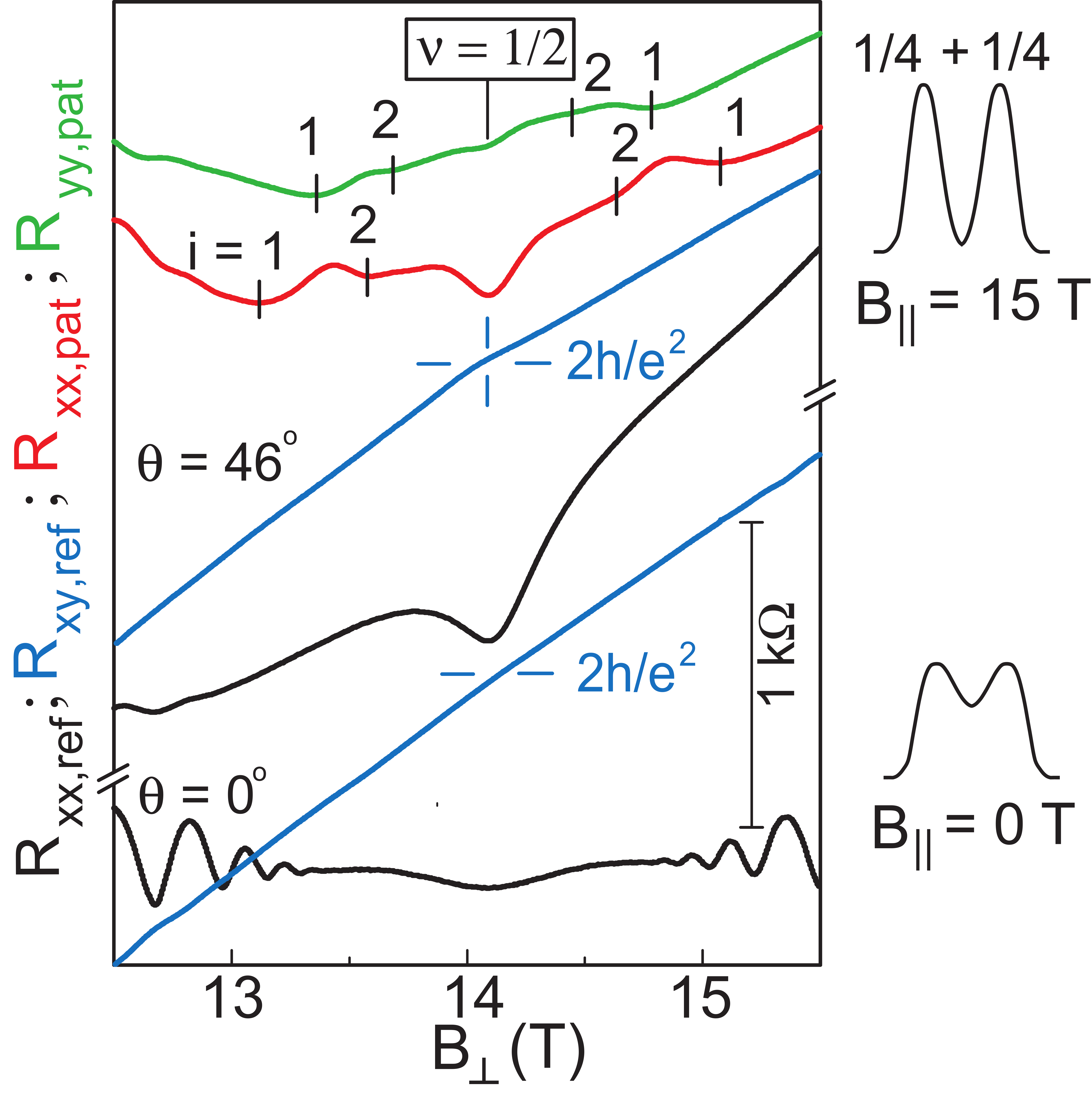}
\centering
\caption{\label{fig:Fig3} (color online) Magnetoresistance data near $\nu=1/2$ at $\theta=0^\circ$ (lower two traces) and $46^\circ$ (upper four traces). The black and blue traces correspond to the reference region while the red and green traces are for the patterned regions.
On the right, we also show the self-consistently calculated (assuming $B_{\perp}=0$) charge distributions for $B_{||}=0$ and 15 T. Traces are shifted vertically for clarity.
}
\end{figure}

Figure \ref{fig:Fig2} illustrates the evolution of the ground-state near $\nu=1$ as a function of $B_{||}$. In Fig. \ref{fig:Fig2}(a), the $\theta=0^{\circ}$ $R_{xx,pat}$ trace (red) shows a wide minimum at $\nu=1$ that reaches zero and no signatures of GR features. (As seen in Fig. \ref{fig:Fig1}, $R_{xx,ref}$ also shows a wide, zero-resistance minimum at $\nu=1$.) We also observe a wide plateau in the $R_{xy,ref}$ trace (blue); these are characteristics of a strong $\nu=1$ IQHS, expected for a single-layer 2DES. In contrast, as $B_{||}$ is increased to $\sim15$ T at $\theta=66^{\circ}$ (Fig. \ref{fig:Fig2}(d)), the $R_{xx,ref}$ trace (black) shows a broad and shallow minimum, superimposed on an increasing background, across $\nu=1$ while no plateau is observed in $R_{xy,ref}$, indicating that the ground-state is now compressible, i.e., the QHS is replaced by a CF Fermi sea. The patterned regions' traces in Fig. \ref{fig:Fig2}(d) show additional resistance minima (marked by vertical lines) on both sides of $\nu=1$. These features, which are absent in the $R_{xx,ref}$ trace, stem from the GR of CFs. 

The $R_{xx,pat}$ and $R_{yy,pat}$ traces in Fig. \ref{fig:Fig2}(d), measured from the perpendicular arms of the L-shaped Hall bar (see Fig. \ref{fig:Fig10}(c)), probe the CF Fermi wave vectors perpendicular and parallel to $B_{||}$, respectively. In Fig. \ref{fig:Fig2}(d) traces, the $i=1$ and 2 resistance minima in $R_{xx,pat}$ are positioned further from $\nu=1$ compared to $R_{yy,pat}$. This is consistent with the $B_{||}$-induced distortion of the CFs' Fermi contour which shrinks along $B_{||}$ and gets elongated in the perpendicular direction \cite{Kamburov.PRL.2013,Kamburov.PRB.2014}. The positions of these GR features yield $k_F^{*}$ along the two principal axes of the Fermi contour. Assuming an elliptical shape for the distorted Fermi contour of CFs \cite{footnoteB}, the deduced $k_F^{*}$ from Fig. \ref{fig:Fig2}(d) translate to a CF density of $9\times10^{10}$ cm$^{-2}$. This is approximately (to within 5$\%$) equal to \textit{half} of the total carrier density, confirming that the charge distribution near $\nu=1$ is bilayer-like for $B_{||}\simeq15$ T. Therefore, the GR features near $\nu=1$ originate from two parallel layers each at $\nu=1/2$. The charge distribution presented in the Fig. \ref{fig:Fig2}(d) inset also corroborates our observation as it clearly shows two distinct layers at $B_{||}=15$ T, each possessing half of the total density.   

Data taken at slightly smaller values of $\theta$ display more intriguing behavior. As the tunneling is increased by decreasing $B_{||}$ (see the $\theta=63^{\circ}$ traces in Fig. \ref{fig:Fig2}(c)), the ground-state at $\nu=1$ evolves back to a QHS as evidenced by the relatively deep and sharp $R_{xx,ref}$ minimum and the developing $R_{xy,ref}$ plateau. However, we still observe GR features (vertical marks) very near $\nu=1$ in the $R_{xx,pat}$ and $R_{yy,pat}$ traces similar to the $\theta=66^{\circ}$ case. These features too correspond to \textit{half} density, suggesting that the charge distribution is still bilayer-like. Similar behavior is also confirmed in the $\theta=61^{\circ}$ $R_{xx,pat}$ traces of Fig. \ref{fig:Fig2}(b) by the presence of GR features surrounding the very strong QHS at $\nu=1$. These GR features, which are consistent with a bilayer charge distribution in the vicinity of $\nu=1$, imply that the observed QHS at $\nu=1$ for $\theta=63^{\circ}$ and $61^{\circ}$ is also of bilayer origin and is therefore the two-component $\Psi_{111}$ state. 

The formation of a $B_{||}$-induced $\Psi_{111}$ state, close to the $\theta$ at which the $\nu=1$ QHS collapses, is consistent with the density-induced evolution (at $B_{||}=0$) in wide QW systems \cite{Lay.PRB.1994}. As the 2DES density is increased to sufficiently high values to render the charge distribution bilayer-like, the ground-state at $\nu=1$ undergoes a phase transition from QHS to a compressible state. It was argued in Ref. \cite{Lay.PRB.1994} that the $\nu=1$ QHS observed near this phase boundary is not the single-component IQHS but rather the $\Psi_{111}$ state. The temperature dependence of the $\nu=1$ QHS near the boundary, quite different from that of the single-component QHS at lower densities, was considered as the main evidence \cite{Lay.PRB.1994}. The GR features we report here very near $\nu=1$ provide a more direct signature for the $\Psi_{111}$ state by confirming its bilayer characteristics for $\theta=63^{\circ}$ and $61^{\circ}$. 
The results also indicate that, while the system creates an exciton condensate at $\nu=1$ in the form of a $\Psi_{111}$ QHS \cite{Eisenstein.ARCMP.2014}, slightly away from $\nu=1$ it morphs into a bilayer (two-component) Fermi gas of CFs \cite{footnoteD,Liu.PRL.2016}. 

Figure 3 also shows that the QHS at $\nu=1$ gets stronger as $\theta$ is decreased. This is consistent with the increase in interlayer tunneling which is expected to strengthen the $\nu=1$ QHS. In contrast, as $\theta$ is decreased from 66$^{\circ}$, the GR features of CFs near $\nu=1$ become progressively weaker and eventually disappear at $\theta=59^{\circ}$ (data not shown). 
The diminishing amplitude of the GR features suggests that, as interlayer tunneling increases with decreasing $B_{||}$, the two-component CF phase becomes unstable. This observation is in agreement with Ref. \cite{Moller.PRB.2009}. 


Data presented so far offer conclusive evidence that CFs near the two-component $\Psi_{111}$ QHS are also two-component. In contrast, as illustrated in Fig. \ref{fig:Fig3}, similar measurements near $\nu=1/2$ provide a stark difference. From Fig. \ref{fig:Fig1} traces of $R_{xx,ref}$ and $R_{xx,pat}$, we know that the ground-state at and near $\nu=1/2$ for $B_{||}=0$ is a CF Fermi sea. Moreover, the GR features, which are consistent with the total density, confirm the single-component nature of CFs near $\nu=1/2$. To compare with $\nu=1$, we show data (Fig. \ref{fig:Fig3}) for $B_{||}\simeq15$ T; this corresponds to $\theta=46^{\circ}$ for $\nu=1/2$. At this $\theta$, away from $\nu=1/2$, the $R_{xx,pat}$ (red) and $R_{yy,pat}$ (green) traces show pronounced GR features (vertical lines marking $i=1$ and 2). Similar to the $\nu=1$ case, the positions of these features reflect the $B_{||}$-induced distortion in the CF Fermi contour. However, unlike $\nu=1$, the deduced CF density of $1.65\times10^{11}$ cm$^{-2}$ agrees well with the $total$ carrier density (within 4\%) suggesting that the CFs near $\nu=1/2$ stay single-component even up to $B_{||}\sim15$ T. 

In the reference region's traces, at $\theta=46^{\circ}$ we observe a sharp minimum in $R_{xx,ref}$ and a developing plateau in $R_{xy,ref}$ at $\nu=1/2$ (Fig. \ref{fig:Fig3}). Considering the $\theta=0^{\circ}$ $R_{xx,ref}$ and $R_{xy,ref}$ traces, which show none of these features, we conclude that $B_{||}$ has induced a bilayer FQHS at $\nu=1/2$, presumably the two-component $\Psi_{331}$ state \cite{Lay.PRB.1997,Hasdemir.PRB.2015}. Note that, the $\nu=1/2$ FQHS is stronger at lower temperatures as reported in Ref. \cite{Hasdemir.PRB.2015}. 
The data of Fig. \ref{fig:Fig3} establish that, although the ground-state of the electron system at $\nu=1/2$ at $\theta=46^{\circ}$ is a bilayer QHS, slightly away from $\nu=1/2$, CFs still behave as single-layer-like (single-component). Similar observations were made for a 2DES, confined to a 65-nm-wide GaAs QW, with density $\sim1.8\times10^{11}$ cm$^{-2}$ \cite{Mueed1.PRL.2015} at $\theta=0^{\circ}$, as mentioned near the beginning of the manuscript. In that case, the wider well width results in a bilayer-like charge distribution and a $\nu=1/2$ QHS without any $B_{||}$. The consistency of CFs' single-component nature near the bilayer QHS at $\nu=1/2$ (for both the $B_{||}$-induced and the $B_{||}=0$ cases) strengthens our finding that CFs behave differently near the bilayer QHSs at $\nu=1$ and $\nu=1/2$. (The single-component CFs near $\nu = 1/2$ should eventually become two-component under very large $B_{||}$. However, we are unable to detect them in our measurements. At such large values of $B_{||}$, the system prefers a bilayer Wigner crystal phase \cite{Manoharan.PRL.1996,Hasdemir.PRB.2015,Hatke.Ncomm.2015}, manifested by very large resistance values and the disappearance of GR features.)

In conclusion, there is a phase transition into a CF Fermi gas state very near the bilayer QHSs at $\nu=1$ and 1/2. While CFs favor a two-component state near $\nu=1$, they stay single-component near $\nu=1/2$. We note that such two- versus single-component contrast between $\nu=1$ and 1/2 in a bilayer system also extends to the FQHSs observed further away from the GR features. For example, the $\nu=$ 6/5 and 4/5 FQHSs near the $\Psi_{111}$ state are two-component \cite{Manoharan.PRL.1997} while the $\nu=3/5$, 5/9 and 5/11 FQHSs near the $\nu=1/2$ QHS are single-component \cite{Suen.PRL.1994,Hasdemir.PRB.2015}. The GR features of CFs, together with these surrounding FQHSs of similar characteristics, thus suggest that while the bilayer QHS at $\nu=1$ is two-component, the $\nu=1/2$ QHS might be single-component.

\begin{acknowledgments}
We acknowledge support through the NSF (Grants DMR-1305691 and ECCS-1508925) for measurements. We also acknowledge the NSF (Grant MRSEC DMR-1420541), the DOE BES (Grant DE-FG02-00-ER45841), the Gordon and Betty Moore Foundation (Grant GBMF4420), and the Keck Foundation for sample fabrication and characterization. Our measurements were partly performed at the National High Magnetic Field Laboratory (NHMFL), which is supported by the NSF Cooperative Agreement DMR-1157490, by the State of Florida, and by the DOE. We thank S. Hannahs, T. Murphy, A. Suslov, J. Park and G. Jones at NHMFL for technical support. We are also grateful to R. Winkler for providing the charge distribution results, and to J. K. Jain and K. Yang for illuminating discussions. 
\end{acknowledgments}


\begin{thebibliography}{99}
\bibitem{Jain.2007} J. K. Jain, \textit{Composite Fermions}
(Cambridge University Press, New York, 2007).
\bibitem{Shayegan.Flatland.Review.2006} M. Shayegan, ``Flatland Electrons in High Magnetic
Fields," in \textit{High Magnetic Fields: Science and Technology}, Vol. 3, edited by Fritz Herlach and Noboru Miura
(World Scientific, Singapore, 2006), pp. 31-60. [condmat/0505520].
\bibitem{Manoharan.PRL.1996} H. C. Manoharan, Y. W. Suen, M. B. Santos, and M. Shayegan,
Phys. Rev. Lett. \textbf{77}, 1813 (1996).
\bibitem{Hatke.Ncomm.2015} A. T. Hatke, Y. Liu, L. W. Engel, M. Shayegan, L. N. Pfeiffer, K. W. West and K. W. Baldwin, Nature Commun. \textbf{6}, 7071 (2015). 
\bibitem{Halperin.HPA.1983} B. I. Halperin, Helv. Phys. Acta. \textbf{56}, 75 (1983).
\bibitem{Suen.PRL.1992} Y. W. Suen, L. W. Engel, M. B. Santos, M. Shayegan, and D. C. Tsui, Phys. Rev. Lett. \textbf{68}, 1379 (1992).
\bibitem{Eisenstein.PRL.1992} J. P. Eisenstein, G. S. Boebinger, L. N. Pfeiffer, K. W. West, and Song He, Phys. Rev. Lett. \textbf{68}, 1383 (1992).
\bibitem{Murphy.PRL.1994} S. Q. Murphy, J. P. Eisenstein, G. S. Boebinger, L. N. Pfeiffer, and K. W. West,
Phys. Rev. Lett. \textbf{72}, 728 (1994).
\bibitem{Suen.PRL.1994} Y. W. Suen, H. C. Manoharan, X. Ying, M. B. Santos, and M. Shayegan, Phys. Rev. Lett. \textbf{72}, 3405 (1994). 
\bibitem{Greiter.PRB.1992} Martin Greiter, X. G. Wen, and Frank Wilczek, Phys. Rev. B \textbf{46}, 9586 (1992).
\bibitem{Lay.PRB.1994} T. S. Lay, Y. W. Suen, H. C. Manoharan, X. Ying, M. B. Santos, and M. Shayegan, Phys. Rev. B \textbf{50}, 17725(R) (1994).
\bibitem{Moon.PRB.1995} K. Moon, H. Mori, Kun Yang, S. M. Girvin, A. H. MacDonald, L. Zheng, D. Yoshioka, and Shou-Cheng Zhang, Phys. Rev. B \textbf{51}, 5138 (1995).
\bibitem{Girvin.Book} S. M. Girvin and A. H. MacDonald, in \textit{Perspectives in Quantum Hall Effects: Novel Quantum Liquids in Semiconductor Structures}, edited by S. DasSarma and A. Pinczuk (Wiley, New York, 1996), p. 161.
\bibitem{Lay.PRB.1997} T. S. Lay, T. Jungwirth, L. Smr\'{c}ka, and M. Shayegan,
Phys. Rev. B \textbf{56}, R7092(R) (1997).
\bibitem{Spielman.PRL.2000} I. B. Spielman, J. P. Eisenstein, L. N. Pfeiffer, and K. W. West,
Phys. Rev. Lett. \textbf{84}, 5808 (2000).
\bibitem{Kellogg.PRL.2004} M. Kellogg, J. P. Eisenstein, L. N. Pfeiffer and K. W. West, Phys. Rev. Lett. \textbf{93}, 036801 (2004).
\bibitem{Tutuc.PRL.2004} E. Tutuc, M. Shayegan, and D. A. Huse, Phys. Rev. Lett. \textbf{93}, 036802 (2004).
\bibitem{Peterson.PRB.2010} M. R. Peterson, Z. Papi\'{c}, and S. Das Sarma, Phys. Rev. B \textbf{82}, 235312 (2010).
\bibitem{Storni.PRL.2010} M. Storni, R. H. Morf, and S. Das Sarma, Phys. Rev. Lett. \textbf{104}, 076803 (2010).
\bibitem{Biddle.PRB.2013} J. Biddle, M. R. Peterson, and S. Das Sarma, Phys. Rev. B \textbf{87}, 235134 (2013).
\bibitem{Shabani.PRB.2013} J. Shabani, Yang Liu, M. Shayegan, L. N. Pfeiffer, K. W. West, and K. W. Baldwin, Phys. Rev. B \textbf{88}, 245413 (2013).
\bibitem{Liu.PRL.2014} Yang Liu, A. L. Graninger, S. Hasdemir, M. Shayegan, L. N. Pfeiffer, K. W. West, K. W. Baldwin, and R. Winkler, Phys. Rev. Lett. \textbf{112}, 046804 (2014).
\bibitem{Liu.PRB.2014} Yang Liu, S. Hasdemir, D. Kamburov, A. L. Graninger, M. Shayegan, L. N. Pfeiffer, K. W. West, K. W. Baldwin, and R. Winkler, Phys. Rev. B \textbf{89}, 165313 (2014).
\bibitem{Eisenstein.ARCMP.2014} J. P. Eisenstein, Annu. Rev. Condens. Matter Phys. \textbf{5}, 159
(2014).
\bibitem{Hasdemir.PRB.2015} S. Hasdemir, Yang Liu, H. Deng, M. Shayegan, L. N. Pfeiffer, K. W. West, K. W. Baldwin, and R. Winkler,
Phys. Rev. B \textbf{91}, 045113 (2015).
\bibitem{Mueed1.PRL.2015} M. A. Mueed, D. Kamburov, S. Hasdemir, M. Shayegan, L. N. Pfeiffer, K. W. West, and K. W. Baldwin,
Phys. Rev. Lett. \textbf{114}, 236406 (2015).
\bibitem{Liu.PRB.2015} Yang Liu, S. Hasdemir, M. Shayegan, L. N. Pfeiffer, K. W. West, and K. W. Baldwin, Phys. Rev. B \textbf{92}, 195156 (2015).
\bibitem{Thiebaut.PRB.2015} N. Thiebaut, N. Regnault, and M. O. Goerbig, Phys. Rev. B \textbf{92}, 245401 (2015).
\bibitem{footnoteC} For simplicity, we have omitted the Gaussian factors from the wave function.
\bibitem{footnoteA} In the original formulation of the two-component wave functions discussed in Ref. \cite{Halperin.HPA.1983}, the components considered were the spin or valley degree of freedom. Since ``layer''  is typically regarded as a pseudo-spin, the wave function also applies to layer components \cite{Girvin.Book}.
\bibitem{Jain.PRL.1989} J. K. Jain, Phys. Rev. Lett. \textbf{63},
199 (1989).
\bibitem{Halperin.PRB.1993} B. I. Halperin, P. A. Lee, and N. Read,
Phys. Rev. B \textbf{47}, 7312 (1993).
\bibitem{Kamburov.PRL.2014} D. Kamburov, Yang Liu, M. A. Mueed,
M. Shayegan, L. N. Pfeiffer, K. W. West and K. W. Baldwin, Phys. Rev. Lett. \textbf{113}, 196801 (2014).
\bibitem{Barkeshli.PRB.2015} Maissam Barkeshli, Michael Mulligan, and Matthew P. A. Fisher
Phys. Rev. B \textbf{92}, 165125 (2015).
\bibitem{Kachru.PRB.2015} Shamit Kachru, Michael Mulligan, Gonzalo Torroba, and Huajia Wang, Phys. Rev. B \textbf{92}, 235105 (2015).
\bibitem{Balram.PRL.2015} Ajit C. Balram, Csaba T\'{o}ke, and J. K. Jain
Phys. Rev. Lett. \textbf{115}, 186805 (2015).
\bibitem{Son.PRX.2015} Dam Thanh Son
Phys. Rev. X \textbf{5}, 031027 (2015).
\bibitem{Balram.PRB.2016} Ajit C. Balram and J. K. Jain
Phys. Rev. B \textbf{93}, 235152 (2016).
\bibitem{Mulligan.PRB.2016} Michael Mulligan, S. Raghu, and Matthew P. A. Fisher,
Phys. Rev. B \textbf{94}, 075101 (2016).
\bibitem{Wang.PRB.2016} Chong Wang and T. Senthil
Phys. Rev. B \textbf{93}, 085110 (2016).
\bibitem{Geraedts.Science.2016} Scott D. Geraedts, Michael P. Zaletel, Roger S. K. Mong, Max A. Metlitski, Ashvin Vishwanath, Olexei I. Motrunich, Science, \textbf{352}, 6282 (2016).
\bibitem{footnote1}Throughout this paper, we use `` * " to denote the CF parameters.
\bibitem{Willett.PRL.1993} R. L. Willett, R. R. Ruel, K. W. West,
and L. N. Pfeiffer, Phys. Rev. Lett. \textbf{71}, 3846 (1993).
\bibitem{Kang.PRL.1993} W. Kang, H. L. Stormer, L. N. Pfeiffer,
K. W. Baldwin, and K. W. West, Phys. Rev. Lett. \textbf{71}, 3850 (1993).
\bibitem{Goldman.PRL.1994} V. J. Goldman, B. Su, and J. K. Jain,
Phys. Rev. Lett. \textbf{72}, 2065 (1994).
\bibitem{Smet.PRL.1996} J. H. Smet, D. Weiss, R. H. Blick,
G. Lutjering, K. von Klitzing, R. Fleischmann, R. Ketzmerick, T. Geisel, and G. Weimann, Phys. Rev. Lett. \textbf{77}, 2272 (1996).
\bibitem{Smet.PRL.1999}J. H. Smet, S. Jobst, K. von Klitzing,
D. Weiss, W. Wegscheider, and V. Umansky, Phys. Rev. Lett. \textbf{83}, 2620 (1999).
\bibitem{Willett.PRL.1999} R. L. Willett, K. W. West, and
L. N. Pfeiffer, Phys. Rev. Lett. \textbf{83}, 2624 (1999).
\bibitem{Kamburov.PRL.2012} D. Kamburov, M. Shayegan,
L. N. Pfeiffer, K. W. West, and K. W. Baldwin, Phys. Rev. Lett. \textbf{109}, 236401 (2012).
\bibitem{Kamburov.PRL.2013} D. Kamburov, Yang Liu, M. Shayegan,
L. N. Pfeiffer, K. W. West, and K. W. Baldwin, Phys. Rev. Lett. \textbf{110}, 206801(2013).
\bibitem{Kamburov.PRB.2014} D. Kamburov, M. A. Mueed, M. Shayegan,
L. N. Pfeiffer, K. W. West, K. W. Baldwin, J. J. D. Lee, and R. Winkler, Phys. Rev. B \textbf{89}, 085304 (2014).
\bibitem{Mueed.PRL.2015} M. A. Mueed, D. Kamburov, M. Shayegan, L. N. Pfeiffer, K. W. West, K. W. Baldwin, and R. Winkler, Phys. Rev. Lett. \textbf{114}, 236404 (2015).
\bibitem{Manoharan.PRL.1997} H. C. Manoharan, Y. W. Suen, T. S. Lay, M. B. Santos, and M. Shayegan,
Phys. Rev. Lett. \textbf{79}, 2722 (1997).
\bibitem{Footnote.1} In this context, increasing $B_{||}$ is equivalent to increasing the 2DES density which also causes the charge distribution in a wide QW to become progressively more bilayer-like \cite{Manoharan.PRL.1996, Suen.PRL.1994, Lay.PRB.1997}.
\bibitem{Skuras.APL.1997} E. Skuras, A. R. Long, I. A. Larkin, J. H. Davies, and M. C. Holland, Appl. Phys. Lett. \textbf{70}, 871 (1997).
\bibitem{Endo.PRB.2005} A. Endo and Y. Iye, Phys. Rev. B \textbf{72}, 235303 (2005).
\bibitem{Kamburov.PRB.2012} D. Kamburov, H. Shapourian, M. Shayegan, L. N. Pfeiffer, K. W. West, K. W. Baldwin and R. Winkler, Phys. Rev. B \textbf{85}, 121305(R) (2012).
\bibitem{footnote1_1} In view of Ref. \cite{Kamburov.PRL.2014}, we have taken CF density to be equal to the minority carrier density in the lowest Landau level. 
\bibitem{footnoteB} According to Refs. \cite{Kamburov.PRL.2013} and \cite{Kamburov.PRB.2014}, the GR features of CFs near $\nu=1/2$ are consistent with an elliptical distortion of the Fermi contour up to large $B_{||}$.
\bibitem{footnoteD} We add that in GaAs 2D hole systems, there is a crossing of lowest two Landau levels \cite{Liu.PRB.2014,Liu.PRB.2015} which stabilizes QHSs at $\nu=1$ and 1/2. In our measurements near such a $\nu=1$ QHS, we observe GR features of CFs which also correspond to half the total density. For the $\nu=1/2$ case, however, the 2D hole system shows insulating behavior \cite{Liu.PRL.2016}.
\bibitem{Liu.PRL.2016} Yang Liu, S. Hasdemir, L. N. Pfeiffer, K. W. West, K. W. Baldwin, and M. Shayegan, Phys. Rev. Lett. \textbf{117}, 106802 (2016).  
\bibitem{Moller.PRB.2009} G. M\"{o}ller, S. H. Simon, and E. H. Rezayi, Phys. Rev. B \textbf{79}, 125106 (2009).
\end{thebibliography}
\end{document}